# From anti-perovskite to double anti-perovskite: lattice chemistry basis for super-fast transportation of $Li^+$ ions in cubic solid lithium halogen-chalcogenides

Zhuo Wang[a,b] , Hongjie Xu[a,b], Minjie Xuan[a,b] and Guosheng Shao[a,b*]

[a] State Center for International Cooperation on Designer Low-carbon & Environmental Materials (CDLCEM), Zhengzhou University, 100 Kexue Avenue, Zhengzhou 450001, China

[b] Zhengzhou Materials Genome Institute (ZMGI), Zhongyuanzhigu, Xingyang 450100, China

*Corresponding author email: gsshao@zzu.edu.cn

**Abstract:** Using a materials genome approach on the basis of the density functional theory, we have formulated a new class of inorganic electrolytes for fast diffusion of $Li^+$ ions, through fine-tuning of lattice chemistry of anti-perovskite structures. Systematic modelling has been carried out to elaborate the structural stability and ion transportation characteristics in $Li_3AX$ based cubic anti-perovskite, through alloying on the chalcogen lattice site (A) and alternative occupancy of the halogen site (X). In addition to identifying effective ways for reduction of diffusion barriers for $Li^+$ ions in anti-perovskite phases via suitable designation of lattice occupancy, the current theoretical study leads to discovery and synthesis of a new phase with a double-anti-perovskite structure, $Li_6OSI_2$ (or $Li_3O_{0.5}S_{0.5}I$). Such a new compound is of fairly low activation barrier for $Li^+$ diffusion, together with a wide energy band gap to hinder conduction of electrons.

## I. Introduction

Solid-state electrolytes (SSE) for lithium-ion battery (LIB) have attracted enormous attention, owing to their highly superior safety advantage over organic liquid ionic conductors and great potential in offering improved electrochemical capacity with lithium anode.[1-4] Still, extensive substitution of liquid organic electrolytes by SSE is yet desired, unless related technologies are further improved to meet the following key criteria: (a) High $Li^+$ ionic conductivity being greater than 1 $mScm^{-1}$(current technical request for liquid organic electrolytes); (b) Low electronic conductivity to avoid self-discharging in service; (c) Broad working temperature for stable operation from −100 to 300 ℃; (d) being electrochemically compatible to lithium anode,[5] and (e) being light, economical, and environmentally friendly.

As a landmark breakthrough, a new family of SSEs based on multi-component sulfides has been shown to provide significantly increased ionic conductivity, making them potential candidates as fast ion conductor for solid lithium-ion batteries. The discovery of $Li_{10}GeP_2S_{12}$ (LGPS),[6,7] which had high bulk $Li^+$ ions conductivity of over 10 $mScm^{-1}$ at room temperature, demonstrated for the first time that $Li^+$ ionic conductivity in solid electrolytes could be superior to that in organic liquid electrolytes. The efficient conduction pathway for $Li^+$ was along the longer c-axis of the tetragonal phase of $Li_{10}GeP_2S_{12}$, with an activation barrier of 0.22-0.25 eV for $Li^+$ diffusion.[6,7] The motivation in replacing the expensive Ge content led to discovery of a cheaper candidate $Li_{9.54}Si_{1.74}P_{1.44}S_{11.7}Cl_{0.3}$,[8,9] which exhibited an even higher ionic conductivity of about 25 $mScm^{-1}$ at room temperature. However, this latter electrolyte was shown not to be electrochemically stable with the lithium anode, which is considered an issue to hinder the full potential of the lithium metal anode, which has the highest energy density (3,860 $mAhg^{-1}$) and lowest potential (-3.4 V *v.s.* standard hydrogen electrode).[10]

Indeed, interface-related problems are still major huddles hindering the progress in the development of all-solid-state LIBs.[10,11,12] While SSE is to be electrochemically stable at the anode Li/SSE interface,[10,11] adequate concentration difference of $Li^+$ ions across the SSE/cathode interface is yet desired, as it helps overcome the interfacial resistance and thus facilitates $Li^+$ diffusing into the cathode materials.[12] It is therefore pressing to develop various alternative SSEs that are able to accommodate higher concentration of $Li^+$ while offering adequate electrochemical voltage windows for improved stability.

Recently, lithium-rich compounds with the anti-perovskite structure, $Li_3OX$ (X = Cl, Br, etc), were synthesized.[13,14] The $Li_3OX$ crystal contains $Li_6O^{+4}$ octahedral units, which are connected via all six vertices to form a three-dimensional framework with a negative monovalent halogen anion $X^-$ sitting at the center.[14] It has been considered a promising alternative SSE,[15,16,17] since it (a) exhibits good thermodynamic and electrochemical stability; (b) is compatible with lithium metal anode; (c) has a large band gap to hinder electronic conduction; and (d) has $Li^+$ ions occupying 60% atomic positions in $Li_3OCl$ with each $Li^+$ having abundant nearest $Li^+$ sites. Also, its cubic structure offers further advantage in permitting three-dimensionally equivalent diffusion of $Li^+$ ions.

The experimentally measured $Li^+$ conductivities for $Li_3OX$, however, were rather scattered. For example, the anti-perovskite $Li_3OCl$ films deposited by pulsed laser deposition (PLD) [15,16] showed ionic conductivity of 0.2 mS/cm with an activation energy $E_a$= 0.35 eV. Work was also carried out to synthesize $Li_3OCl$ by heating at 360 ℃ under vacuum for several days.[17] The measured $Li^+$ ionic conductivity for the latter was 0.85 mS/cm together with an activation energy of 0.26 eV at room temperature.

Some theoretical efforts were attempted in order to account for the scattered experimental data from the $Li_3OX$ alloys of interest.[18-20] It was suggested that there are two manners for $Li^+$ diffusion in $Li_3OCl$ , either by vacancy[18-20] or interstitial $Li^+$.[20] The first one was realized by $Li^+$ migration into a neighboring vacant site of its kind. The activation energy along this pathway was determined to be in the range from 0.31 to 0.37 eV, by the nudged elastic band (NEB) method or from *ab initio* molecular dynamics (AIMD) simulation.[18-20] Such theoretically predicted activation energies are in good agreement with the experimentally measured value of 0.35 eV.[15,16] The second mechanism was fulfilled via migrating an interstitial $Li^+$ to a neighboring site of its kind through an interstitial channel in the form of a dumbbell-like pathway. It was suggested that such an interstitial pathway would enable ultrafast $Li^+$ diffusing channel with the activation energies in the range from 0.145 to 0.175 eV.[20] This indicated that moderate variation of $Li^+$ concentration in the SSE could have significant impact on the $Li^+$ transportation. One notices that the ionic conductivity achieved in $Li_3OCl$ was still far inferior to that from the LGPS family (0.85 vs. 10 mS/cm). Subsequent experimental work showed that slight deviation from the standard $Li_3OCl$ stoichiometry resulted in a glassy $Li_{2.99}Ba_{0.005}O_{1+x}Cl_{1-2x}$ alloy with significantly improved ionic transport properties, with $Li^+$ conductivity ($\sigma$) being over 10 mS/cm and a negligible activation energy of 0.06 eV.[21] It was postulated that the significant improvement in performance was because of the absence of anode-electrolyte interphase (SEI), with a glassy type of electrolyte discouraging nucleation of SEI crystals.

To date, systematic theoretical efforts are yet desirable to guide materials design for ultrafast $Li^+$ conductivity, through fine-tuning of alloy chemistry on the basis of the anti-perovskite structure. Also, the $Li_3OCl$ phase tends to decompose into more stable LiCl and $Li_2O$ phases at service temperature of interest for LIBs.[22,23] As the $Li_2O$ phase is a rather poor ionic conductor, it would be useful in examining the feasibility in substitution of O by other chalcogen elements such as S. It is recognized that $Li_2S$ is a much better ionic conductor, than $Li_2O$.[24-26]

Here in this work, we aim to identify effective ways for radical reduction of diffusion barriers for $Li^+$ ions in anti-perovskite phases $Li_3AX$ (chalcogen A, halogen X), through either suitable chemical deviation from the standard stoichiometry of the tri-lithium halogen-chalcogenides, or by variation of the lattice chemistry associated with the A and X sites. We show that very low activation energy for $Li^+$ diffusion can be achieved with the combined effect of Li and A enrichment and X deficiency. A new double anti-perovskite compound with the stoichiometry of $Li_6OSI_2$ is identified theoretically and then successfully synthesized. Such a double anti-perovskite phase is thermodynamically stable, with ionic conductivity rivalling that for the well-known LGPS sulfide. Also, suitable tuning in lattice chemistry off the stoichiometric composition is shown to result in significantly improved ionic transportation, with an activation of 0.18 eV to enable more superior $Li^+$ conductivity well below the room temperature. The outcome of this work is expected to open a new avenue in synthesis of novel Li-rich halogen-chalcogenide SSE suitable for all-solid LIBs.

## II. Method

Theoretical calculations are performed using the Vienna *Ab initio* Simulation Package (VASP),[28,29] with the ionic potentials including the effect of core electrons being described by the projector augmented wave (PAW) method.[30,31] In this work, the Perdew−Burke−Ernzerhof (PBE) GGA exchange−correlation (XC) functionals[32,33] are used to study the structural stabilities in the $Li_3AX$ family. For the geometric relaxation of the structures, summation over the Brillouin Zone (BZ) is performed with 3×3×3 and 5×5×5 Monkhorst−Pack k-point mesh for the conventional and primitive cells respectively. A plane-wave energy cutoff of 600 eV is used in all calculations. All structures are geometrically relaxed until the total force on each ion was reduced below 0.01 eV/Å. For the calculations of electronic energy band structures, we use the HSE06 functional to predict more accurate values of band gaps.[34,35] We employ a convergence criterion of $10^{-6}$ eV for electronic self-consistent cycles.

The climb image nudged elastic band (CI-NEB) method with the Limited-memory Broyden-Fletcher-Goldfarb-Shanno (LBFGS) optimizer[36,37] has been used to search the $Li^+$ diffusion channel in the electrolytes of interest. The initial and final configurations are obtained after full structural relaxation. The number of inserted images used in the CI-NEB calculations depend on the reaction coordinates between the initial and final configurations. This method is useful not only in estimating the activation barrier but also in locating the bottleneck along the $Li^+$ transporting pathway.

In addition, we also carry out *ab initio* molecular dynamics (AIMD) simulations using system units containing 40 atoms and a 2 fs time step in the NVT ensembles with a constant volume and with a Nosé–Hoover thermostat, as a complimentary approach to the CI-NEB method for deriving full diffusion coefficient while checking the activation energy from the statistical point of view.[18] Each AIMD simulation lasts for 80 ps after a 10 ps pre-equilibrium run. In order to shorten the simulation time, elevated temperatures from 750 K up to 2000 K are applied to speed up the ion-hopping process.

The universal structure predictor (USPEX)[38,39] based on a materials genome algorithm is employed to predict stable or metastable structures for given compositions. For each composition, a population of 200 possible structures is created randomly with varied symmetries in the first generation. When the full structure relaxation is reached, the most stable and metastable structures, through the comparison of enthalpy of formation, will be inherited into the next generation. Afterwards, each subsequent generation will be created through heredity, with lattice mutation and permutation operators being applied and assessed energetically for the selection of a population of 60 for next run. USPEX will continue screening the structures until the most stable configuration keeps unchanged for further 20 generations to safeguard global equilibrium.

The phonon frequency spectrum of a theoretically predicted structure is used for examining its dynamical stability. The super-cell method in the PHONOPY package[40,41] is employed to perform the relevant frozen-phonon calculations based on harmonic approximation. The supercells on the basis of the relaxed structures are used for the phonon calculations. The stability criterion is that the amplitude of imaginary frequency is less than 0.3 THz,[42-43] to accommodate numerical errors in phonon calculations.

## III. Results and Discussion

***Global searching for energetically preferred structural configurations for each given composition:***

We start from identifying the stable/metastable structures in each $Li_3AX$ phase through isovalent replacement of either the chalcogen A site or the halogen X site, using the USPEX method for global energy minimization of each chemical configuration.[38,39] Such a theoretical approach is particularly useful when little is known about phase structures in a new material system to be formulated, so that potential phase structures can be predicted with associated properties simulated at 0 K. **Table 1** summarizes the USPEX search results for stable structures with geometrically relaxed lattice parameters for compositions covering $Li_3OCl$, $Li_3OI$, and $Li_6OSI$, which have thermodynamically preferred structures based on the anti-perovskite phase ($Pm\bar{3}m$). One notes that complete substitution of Cl with I in the $Li_3OCl$ perovskite phase leads to some moderate increase in lattice parameter. Complete replacement of O by S in the $Li_3OI$ phase, on the other hand, gives rise to significant increase in lattice parameter together with considerable lattice distortion off the perovskite symmetry. It is amazing to note that the stable structure gains higher symmetry when half of the O site is replaced by S, with $Li_6OSI_2$ having a face-centered cubic space group ($Fm\bar{3}m$) instead ($Pm\bar{3}m$).

**Fig. 1** compares the energies of formation for each stable/metastable configuration at the ground state. Here we follow the convention to formulate the energy of formation with respect to the chemical potentials of stable structures for constituent elements. We can find that binding is stronger when the chalcogen site is occupied by O than S, while Cl occupancy of the halogen site leads to more stable structure than it being occupied by I, $OH^{-1}$, or $NH_2^{-1}$. Apparently, weakened binding due to A or X substitution leads to increased lattice parameters.

The stable structure for the $Li_3OCl$ composition is of the anti-perovskite lattice, as is shown in **Fig.2 (a)**, where $Li_6O^{+4}$ octahedral unit on the corners of a cubic frame, leaving $Cl^-$ sitting at the body center thus leading to an overall lattice symmetry of $Pm\bar{3}m$ *(221)*. However, such a cubic structure is not energetically favored once the chalcogen site is replaced by S, with the $Li_3SCl$ tends to adopt a layered structure, **Fig.2 (b)**. Such a layered structure is typical for the tri-lithium chlorine-sulfide when $Cl^{-1}$ is further replaced by $Br^{-1}$ or a non-halogen function groups such as $NH_2^{-1}$ of $OH^{-1}$. Owing to the significant increase in lattice parameter associated with weakened binding by substituting O by S, it is reasonable to reckon that the ionic radius of $Cl^-$ is too small to support the framework of $Li_6S$ octahedrons, thus leading to collapse of an anti-perovskite structure. Indeed, when a bigger halogen ion $I^-$ is used to replace $Cl^{-1}$, the perovskite structure is largely maintained, **Fig. 2(c)**, albeit experiencing some lattice distortion (**Table 1**). This indicates the applicability of a tolerance factor such as the well-known Goldschmidt rule that dictates the maintenance of the cubic anti-perovskite structure.[27]

The Goldschmidt factor for ionic radius tolerance factor for an anti-perovskite phase $Li_3AX$ can be defined as

$$\boldsymbol{t} = \frac{(\boldsymbol{R_{Li}}+\boldsymbol{R_X})}{\sqrt{2}(\boldsymbol{R_{Li}}+\boldsymbol{R_A})},$$

where ***t*** should stay in the range from 0.8 to 1. The tolerance factors for $Li_3SX$ (X= $OH^-$, $NH_2^-$, $Cl^-$, $Br^-$) is then in the range from 0.58 to 0.74 (**Table 2**), being smaller than the lower limit of 0.8. Indeed, they are in accord with the outcome from global USPEX searching, with all of them having layered stable structures. In order to increase the tolerance factor of this family, obviously we should choose a negative monovalent anion with larger ionic radius to occupy the X site. For example, when the X site is occupied by $I^-$, the tolerance factor of $Li_3SI$ is increased to 0.81, leading to a stable structure for $Li_3SI$ being fairly close to the anti-perovskite phase with marginal lattice distortion. On the other hand, when the chalcogen site is occupied by $O^{-2}$, the smaller $O^{-2}$ radius of 1.4 Å requests that the X site needs to be filled by smaller anions to meet the tolerance factor to sustain a cubic structure, so that the anti-perovskite phase is stabilized in the $Li_3OCl$ (0.85), $Li_3OBr$ (0.9), and $Li_3OI$ (0.95) phases (with corresponding tolerance factors in the brackets being all considerably larger than 0.8).

As both $Li_3OI$ and $Li_3SI$ are of anti-perovskite type structures (**Fig. 2(a)** vs. **Fig. 2(c)**), one would be encouraged to mix $O^{-2}$ and $S^{-2}$ in the chalcogen site. The global energy minimization from the USPEX process leads to a double anti-perovskite structure, which is constructed using alternating $Li_6O$ and $Li_6S$ octahedrons over a face-centered cubic lattice, $Li_6OSI_2$ as shown in **Fig.2 (d)**. Such a face centered structure owing to alternating arrangement of the $Li_6O$ and $Li_6S$ units from two anti-perovskite lattices is thus referred as a *double anti-perovskite phase* (in the same spirit to define the double perovskite structure in the literature). The relaxed half lattice parameter equals to 4.3 Å (8.6/2), being slightly smaller than the average from 4.16 Å for $Li_3OI$ and 4.735 Å for $Li_3SI$. Such negative deviation from the linear average of the lattice parameters for the constituent phases, 4.44 Å, results from enhanced binding from alloying on the chalcogen site. The weighted sum of the tolerance factor of the constituent phases (0.81 and 0.98) is well over the 0.8 criterion, which is in accord with the stable cubic structure in the mixed phase.

***Dynamic stability:***

Dynamical stabilities of the fully relaxed structures of energetically stable configurations at 0 K from USPEX searching are checked through examining the characteristics of their phonon band structures. A phase is dynamically stable when no phonon bands are associated with imaginary frequencies, and such a phase is then considered more likely to exist in nature owing to safeguarded mechanical stability. **Fig. 3(a)** to **(d)** are calculated phonon band structures for $2\times2\times2$ supercell of (a) $Li_3OCl$ (40 atoms), (b) $Li_3OI$, and (c) $Li_3SI$ correspondingly. The primitive cell for $Li_6OSI_2$ (d) is based on a face-centered cubic space group ($Fm\bar{3}m$) having ten atoms. Thus, for $2\times2\times2$ supercell of $Li_6OSI_2$, it contains 80 atoms. The absence of imaginary frequencies from the phonon band structures of $Li_3OCl$ and $Li_3OI$ configurations correspond to their big tolerance factors to sustain dynamically stable structures without noticeable lattice distortion. On the other hand, in the phonon band structure for $Li_3SI$, there is some imaginary frequencies below -0.3 THz border line, which is also in accord with the marginal value of the tolerance factor and associated slight lattice distortion from the cubic lattice, albeit the degree of dynamical instability being rather trivial. When the $Li_3SI$ is alloyed with the much more stable $Li_3OI$ into a double anti-perovskite phase of $Li_6OSI_2$, its significantly bigger tolerance factor of 0.895 then leads to a mechanically stable phase, **Fig. 3(d)**.

***Entropy effect on phase equilibria:***

Let us move on to check the stability of the above identified ternary or quaternary compounds with respect to constituent binary phases. We can construct pseudo-ternary phase diagrams of interest, on the basis of energy of formation, with reference to their potential constituent binary phases such as those collected in the Materials Project (MP).[44] Fig. 4 shows pseudo-ternary phase diagrams to contain the $Li_3OCl$, $Li_3SI$, and $Li_6OSI_2$ phases correspondingly. None of them is stable with respect to its constituent phases at 0 K. This is to say that due to their energies of formation being less negative than the linear combinations of their respective constituent phases, energetically they tend to decompose without consideration of any thermal effect:

$$Li_3OCl \rightarrow Li_2O + LiCl,$$

$$Li_3SI \rightarrow Li_2S + LiI,$$

$$Li_6OSI_2 \rightarrow Li_2O + Li_2S + 2LiI,$$

We now examine the thermal effects on free energies of compounds within the quasi-harmonic approximation to consider the phonon entropy.[22,45] The free energies for $Li_3OCl$, $Li_3SI$, and $Li_6OSI_2$ can be calculated against stable constituent compounds by,

$$\Delta G_{\mathrm{Li3OCl}} = G_{\mathrm{Li3OCl}} - G_{\mathrm{Li2O}} - G_{\mathrm{LiCl}},$$

$$\Delta G_{\mathrm{Li3SI}} = G_{\mathrm{Li3SI}} - G_{\mathrm{Li2S}} - G_{\mathrm{LiI}},$$

$$\Delta G_{\mathrm{Li6OSI2}} = G_{\mathrm{Li6OSI2}} - G_{\mathrm{Li2O}} - G_{\mathrm{Li2S}} - 2G_{\mathrm{LiI}},$$

where all ground state stable constituents such as $Li_2O$, $Li_2S$, LiCl, and LiI are found to be also dynamically/mechanically stable. The corresponding energy changes are termed as the energy above Hull in literature, with a positive energy above the Hull being indicative of thermodynamic tendency for decomposition into more stable constituent phases.[44]

The critical temperatures to enable negative $\Delta G$ for the cubic phases $Li_3OCl$, $Li_3SI$, and $Li_6OSI_2$ are indicated in **Fig. 5**. It suggests that the cubic $Li_3OCl$ will be more stable than the combination of $Li_2O$ and LiCl above 538 K, which is below its experimentally measured melting point of 282 ºC (555 K),[17]and thus explains why the phase was successfully synthesized at elevated temperatures. The critical temperature to stabilize $Li_3SI$ is about the same (520 K). To compare, the critical temperature to thermally stabilize the $Li_6OSI_2$ is significantly lower than those for $Li_3OCl$ and $Li_3SI$, which is only 145 ºC (418 K). The critical temperatures for thermal stabilization of $Li_6OSI_2$ and $Li_3SI$ are reasonably low, which is helpful for their synthesis. A moderate critical temperature also helps to hinder their decomposition below the critical temperature, due to limited degree of long-range diffusion of all chemical species other than $Li^+$. Indeed, the newly discovered double anti-perovskite phase $Li_6OSI_2$ has been successfully synthesized within our team under the guidance of the current theoretical work. The experimental results are briefly summarized in the supplementary material, with the X-ray diffraction pattern from the new phase being in excellent agreement with the simulated diffraction pattern, **Fig. s1**. Further work is underway to make use of the new SSE in LIBs.

***Electrochemical compatibility with Li anode:***

The electrochemically compatible window with the anode Li metal can be examined according to interfacial reactions against Li uptake per formula unit (f.u.) of solid electrolyte.[46-48] The average electrochemical potential, $\bar{V}_{\mathrm{A}\rightarrow\mathrm{B}}$, for the transition between state A ($Li_x\prod$) and state B ($Li_{x+\Delta x}\prod$), with reference to electrochemical potential for the lithium metal is related to total energies ($E_{total}$) as,

$$\bar{V}_{\mathrm{A}\rightarrow\mathrm{B}} = -1/z \cdot \{\frac{[E_{total}(Li_{x+\Delta x}\prod) - E_{total}(Li_x\prod)]}{\Delta x} - E_{total}(Li)\}$$

Where $x$ is the number of Li in the formula unit of $Li_x\prod$, charge value z = 1 for $Li^+$, and $\Delta x$ is the change in the number of Li atoms, and $\prod$ refers to the collection of other constituents. Again, energetically stable constituent phase structures for each Li level are identified from the USPEX global searching.

Figure 6 (a) and (b) plots the voltage above Li/$Li^+$ against the Li uptake per formula unit of $Li_3OI$ and $Li_3SI$, respectively. The solid electrolyte is to undergo reduction and uptake of Li next to the alkali metal anode (Li rich side) at low voltage, while at high voltage (Li poor side), electrolyte will be oxidized to deplete Li. The intrinsic electrochemical window for the $Li_3OI$ composition corresponds to a range of voltage (0.0−2.4 V), while the width of electrochemical window is between 0.0 and 2.15 V for $Li_3SI$. On the basis of the electrochemical window according to the associated equilibria phases at 0 K, the upper limit of electrochemical window for $Li_6OSI_2$ should be 2.15 V, in which both $Li_3OI$ and $Li_3SI$ will be electrochemically stable.

***$Li^+$ diffusion along vacancy and interstitial $Li^+$ transporting pathways:***

Starting from Li transportation in anti-perovskite structures, diffusion path associated with $Li^+$ vacancies occurs with a $Li^+$ ion migrating into a nearest vacancy site of its kind, thus leaving a vacancy behind. The process for such a migrating mechanism is identified from CI-NEB, as is illustrated in the lower panel images of **Figure 7 (a)**, which captures the typical states for a $Li^+$ diffusing into a neighbor vacancy along the edge of an S centered octahedral unit ($Li_6S$) in $Li_3SI$. The activation energy barrier is 0.27 eV (**Table 3**), which is very close to that in a typical LGPS electrolyte.[6,7]

For the migration of an interstitial $Li^+$ ion in an anti-perovskite phase such as $Li_3SI$, structural relaxation reveals that the presence of an interstitial $Li^+$ tends to offset a vertex $Li^+$ of an $Li_6S$ octahedron from its ideal position, thus forming a dumbbell coordination with the approaching interstitial $Li^+$ at the vertex point. Such a dumbbell formation-decomposition was identified by Emly et al. as the dictating mechanism for the migration of interstitial $Li^+$ ions.[19] The dumbbell-linked pathway for the transportation of an interstitial $Li^+$ is depicted in the lower panel images of Fig. 7(b). Initially, the first interstitial $Li^+$ (1) offsets its vertex neighbor into the second position (2) to form a dumbbell-like microscopic configuration (1-2), while the third $Li^+$ stays at another vertex site. It follows that the first $Li^+$ then fills the previous vertex site of $Li^+$(2) when the latter offset $Li^+$(3) to form another dumbbell coordination (2-3). The activation energy barrier along such a pathway is 0.16 eV, which is about one-third lower than that associated with the vacancy mechanism.

Similarly, $Li^+$ diffusing along vacancy and dumbbell pathways in the $Li_3OCl$ compound are also calculated. Along the vacancy pathway, the activation barrier is 0.35 eV for $Li_3OCl$, which is in excellent agreement with reported value from NEB or AIMD simulation [18-20] and experimental value of 0.35 eV.[15,16] Such an activation barrier is one-quarter higher than the 0.27 eV barrier in $Li_3SI$. Along the dumbbell pathway in $Li_3OCl$, the energy barrier is 0.162 eV, that agrees well with previously calculated value of 0.17 eV.[19]

The charge density distributions for the mid-path sections along the vacancy pathway in $Li_3OCl$ and $Li_3SI$ are compared in **Fig. 8 (a, b)**. The charge maps correspond to the blue-dashed line linked frameworks as the bottlenecks for a $Li^+$ ion to squeeze through. The triangular bottleneck gate constructed by Cl-O-Cl and I-S-I are considerably different in sizes, with S replacement of O leading to a larger gate, making it easier for Li to go through in the latter. The lower value of electronegativity of S than that of O is behind the weaker bonding associated with S substitution of O. From **Fig. 8(a)**, obvious charge overlap between $Cl^-$ and $Li^+$ can be found, while there is little charge overlap between $I^-$ and $Li^+$ in **Fig. 8 (b)**. Therefore, the weaker bonding in $Li_3SI$ leads to lower activation barrier for $Li^+$ diffusion than in $Li_3OCl$.

As is shown in **Fig. 8 (c, d)**, when an excessive $Li^+$ is incorporated into the systems, the bonding length of Li-O increases from 1.77 Å to 1.91 Å, while the bonding length of Li-S increases from 2.26 Å to 2.45 Å. The enlarged lengths between $Li^+$ and the chalcogen anions suggest that the interactions between them are further weakened owning to the presence of excessive $Li^+$ in the systems. This makes it easier for $Li^+$ migration, and the formation of coordinated dumbbell next to the associated vertex site is effective in reducing anion-cation interaction further due to diluted electrostatic interaction. Such combined effects lead to significantly reduced activation barrier for the migration of interstitial $Li^+$ ions.

The double anti-perovskite $Li_6OSI_2$ identified in this work is made of alternating $Li_6O$ and $Li_6S$ octahedrons together with halogen $I^-$ occupying the center of each cubic cage of the octahedrons. Each vertex $Li^+$ is therefore shared by both $Li_6O$ and $LI_6S$ octahedrons, though the Li-chalcogen bond lengths are rather different in the different types of octahedrons. Naturally, the activation barriers associated with either the vacancy or interstitial $Li^+$ diffusing pathways will be impacted by such two kinds of octahedrons. Taking the dumbbell mechanism for interstitial $Li^+$ for example, there are two types of migrating modes associated with the two types of octahedrons, as shown in **Fig. 9**. While the activation barrier for a dumbbell unit to transport around $Li_6S$ is slightly larger than that in $Li_3SI$ and $Li_3OCl$ (0.166 eV vs. 0.16 eV), the activation barrier associated with the $Li_6O$ octahedron is somewhat lower, 0.107 eV.

***The effect of off-stoichiometric components on $Li^+$ transportation:***

Recently, Deng et al. reported that in the cubic argyrodite phase of $Li_6PS_5Cl$, slight off-stoichiometric composition such as $Li_{6.25}PS_{5.25}Cl_{0.75}$ brings about significant enhancement in ionic conductivity.[49] Such effect is also observed to hold in the current alloys, with extraneous $Li^+$ together with chalcogen substitution to halogen to maintain overall charge neutrality. The energetic cost for such slight off-stoichiometric drift can be rather trivial. For example, with some slight deviation from the $Li_6OSI_2$ ($Li_{24}O_4S_4I_8$) stoichiometry, the formation energy for $Li_{25}O_4S_5I_7$ ($Li_{24+1}O_4S_{4+1}I_{8-1}$) is only 0.014 eV per atom above that for $Li_6OSI_2$, suggesting a high likelihood to realize in experiments. The diffusion activation barrier for Li for the former is reduced to only 0.026 and 0.053 eV for the dumbbell pathways around the $Li_6S$ and $Li_6O$ octahedrons respectively (**Table 3**).

***AIMD simulation:***

Although the CI-NEB method can be used to assess the activation barriers for ionic diffusion through path searching over a static energy landscape at 0 K, it is desirable to assess the temperature dependence of the ionic diffusion coefficient *D*. Recent work has demonstrated successful assessment of *D* by AIMD simulation at elevated temperatures from a statistics point of view.[45,49,50] From an AIMD simulation, the diffusion coefficient *D* can be extracted via the following equation:

$$D = \frac{1}{2dt}\langle[\Delta r(t)]^2\rangle,$$

where the dimensionality factor *d* equals to 3 for three-dimensional structures, and $\langle[\Delta r(t)]^2\rangle$ is the average mean square displacement (MSD) over a time duration t. Correction of the MSD from artefactual errors due to the periodic boundary condition is necessary for dependable AIMD evaluation of *D*, using the "unwrapped" trajectories to help achieve significant improvement in sampling statistics from limited simulation data typical for tractable AIMD runs.[51]

On the basis of the Arrhenius relation, the diffusion coefficient, $D = D_0 \exp\left(-\frac{E_a}{k_B T}\right)$, can then be derived through simulating $D$ at a series of temperatures, $T$ ($E_a$ is the activation barrier for $Li^+$ ion transport, $k_B$ is the Boltzmann constant, and $D_0$ is a constant). **Fig. 10** plots the Arrhenius relationships by the logarithmic value of $D$ versus $1/T$, with linear fitting resulting in the activation energies and diffusion coefficient $D$. The diffusion data for various alloys are listed in **Table 3** to compare with data obtained from the CI-NEB method and reported data from the literature.

In the case of the vacancy mechanism, the AIMD values of activation barriers for $Li^+$ transport agree well with the CI-NEB results. On the other hand, the AIMD activation energies associated with interstitial $Li^+$ are considerably larger than those from the dumbbell mechanism, indicating its lesser role for bulk ionic transport. The complexity for coordinated migration in the dumbbell association hinders their occurrence due to lowered probability in a statistical way typical for molecular dynamics. AIMD data are therefore preferred for self-consistent assessment of ionic transport in the alloys of interest in this work.

The alkali ionic conductivity at room temperature $\sigma$ can be derived by the Nernst−Einstein equation as the product of the volume ion density ρ, the square of the ionic charge z, and the absolute mobility of the ion: [52]

$$\sigma = \rho z^2 D / k_B T,$$

Where $\rho$ is the molar density of diffusing alkali ions in the unit cell, z is the charge of alkali ions (+q), and R are the Faraday's constant and the gas constant, respectively. This leads to $Li^+$ conductivity at 300 K for $Li_6OSI_2$ (5 $mScm^{-1}$) and $Li_{25}O_4S_5I_7$ (12.5 $mScm^{-1}$), which rivals the conductivity in current liquid organic electrolytes for LIBs. The lower activation energies achievable in the newly identified double-anti-perovskite phase also makes them more attractive than the well-known LPGS system below room temperature.

***Bandgap:***

While SSE should be highly conductive to $Li^+$ ions, it should be insulating for the conduction of electrons to avoid internal discharging. **Fig. 11** summarizes the density of states (DOS) for the ionic conductors of interest to this work, using the HSE06 screened hybrid functional for improved accuracy in band structure calculation, by accounting for some non-local effect in the exchange-correlation functional. It can be seen that all of the materials are insulating to electrons due to the existence of rather large forbidden energy gaps between the valence band maximum (VBM) and conduction band minimum (CBM). Both the VBM and CBM are dominated by electron states of anions in all compounds. Moreover, the forbidden gap is shown to be mainly determined by the type of chalcogen species (O, S), so that the gap value is correlated to the overall bonding strength. The most strongly bonded O containing material has the largest forbidden gap than the S based materials, $Li_3OCl$ (6.2eV)>$Li_3OI$ (5.3eV)>$Li_6OSI_2$ (4.5eV)> $Li_3SI$ (4.43eV). The charge neutrality in $Li_{25}O_4S_5I_7$ is shown to safeguarding its intrinsic nature as a wide gap material, without any defect states presenting in the forbidden gap, **Fig. S2**.

## IV. Conclusions

Systematic modeling has been carried out to design fast $Li^+$ ion conductors based on the anti-perovskite structure. This enables discovery of two new phases, the anti-perovskite $Li_3SI$ and the double anti-perovskite $Li_6OSI_2$. The latter is especially attractive owing to combined energetic and dynamic stability. Both new phases are of attractively low activation energies for the diffusion of $Li^+$ ions.

Moderate off-stoichiometric compositional deviation in the double-anti-perovskite materials can lead to significant reduction of the activation barrier for $Li^+$ transportation. This includes moderate enrichment of Li ions together with chalcogen excess and halogen deficiency to maintain charge neutrality. The combined effect can lead to low activation energies below 0.18 eV for the double anti-perovskite phase, making it particularly attractive for low temperature applications in whole-solid LIBs.

## Acknowledgement

This work is supported in part by the 1000 Talents Program of China, the Zhengzhou Materials Genome Institute, the National Natural Science Foundation of China (Nos.51001091, 111174256, 91233101, 51602094, 11274100), and the Fundamental Research Program from the Ministry of Science and Technology of China (no. 2014CB931704).

**Table 1** Lattice parameters and symmetry groups for $Li_3OCl$, $Li_3SI$, and $Li_6OSI_2$.

| | a | b | c (Å) | alpha | beta | gamma(°) | Symmetry |
|---|---|---|---|---|---|---|---|
| **$Li_3OCl$** | 3.89 | 3.89 | 3.89 | 90 | 90 | 90 | $Pm\bar{3}m$ *(221)* |
| **$Li_3OI$** | 4.16 | 4.16 | 4.16 | 90 | 90 | 90 | $Pm\bar{3}m$ *(221)* |
| **$Li_3SI$** | 4.735 | 4.748 | 4.738 | 90.03 | 89.99 | 90 | *P1* |
| **$Li_6OSI_2$** | 8.60 | 8.60 | 8.60 | 90 | 90 | 90 | $Fm\bar{3}m$ *(225)* |
| **--(primitive-cell)** | 6.08 | 6.08 | 6.08 | 60 | 60 | 60 | |

**Table 2** Tolerance factor from the Goldschmidt rule

| Ions and Ionic Radius | Tolerance Factor |
|---|---|
| Li (0.68), O (1.4), Cl (1.81) | **0.85** |
| Li (0.68), O (1.4), Br (1.96) | **0.9** |
| Li (0.68), O (1.4), I (2.2) | **0.98** |
| | |
| Li (0.68), S (1.84), Cl (1.81) | 0.7 |
| Li (0.68), S (1.84), Br (1.96) | 0.74 |
| Li (0.68), S (1.84), I (2.2) | **0.81** |
| Li (0.68), S (1.84), $CH_3$ (2.6) | **0.92** |
| Li (0.68), S (1.84), OH (1.40) | 0.58 |
| Li (0.68), S (1.84), $NH_2$ (1.71) | 0.67 |
| | |
| {Li (0.68), O (1.4), I (2.2),<br>Li (0.68), S (1.84), I (2.2)} | **{0.98,<br>0.81} → 0.895** |

**Table 3** Summary of data for ionic diffusion and conductivity, with reference to data from literature. Data from this work is emboldened.

| System | Ea (eV) | ρ ($10^{23}$ $cm^{-3}$) | $D_{300K}$($cm^2/s$) | σ(mS $cm^{-1}$) 300K | σ(mS $cm^{-1}$) 223K |
|---|---|---|---|---|---|
| $Li_3OCl$ vac | *0.35*[exp15,16] | | | *0.2*[exp16] | |
| | *0.303*[AIMD18] | | | *0.12*[AIMD18] | |
| | *0.34*[CINEB19] | | | | |
| | **0.35**[CINEB] | | | | |
| | **0.31**[AIMD] | **0.488** | **$3.23\times10^{-10}$** | **0.1**[AIMD] | |
| $Li_3OCl$ int | *0.26*[exp17] | | | *0.85*[exp17] | |
| | *0.17*[CINEB19] | | | | |
| | **0.162**[CINEB] | | | | |
| | **0.27**[AIMD] | **0.530** | **$3.53\times10^{-9}$** | **1.1**[AIMD] | |
| **$Li_3SI$ vac** | **0.27**[CINEB] | | | | |
| **$Li_3SI$ int** | **0.16**[CINEB] | | | | |
| **$Li_6OSI_2$ int** | **0.107, 0.16**[CINEB] | | | | |
| | **0.22**[AIMD] | **0.393** | **$2.07\times10^{-8}$** | **5.0**[AIMD] | |
| **$Li_{25}O_4S_5I_7$** | **0.026, 0.053**[CINEB] | | | | |
| | **0.18**[AIMD] | **0.393** | **$5.11\times10^{-8}$** | **12.5**[AIMD] | **1.1**[AIMD] |
| LGPS | *0.22*[exp6,7] | | | *10*[exp6,7] | |
| | *0.21*[AIMD50] | | | *9*[AIMD50] | |
| | **0.23**[AIMD] | **0.205** | **$7.63\times10^{-8}$** | **9.7**[AIMD] | **0.5**[AIMD] |

**Fig. 1** The global USPEX searched energies $Li_3AB$ various alloys, while Layered (black), Anti-perovskite (blue), and Anti-perovskite-like (green) structures are obtained at variable chemical compositions.

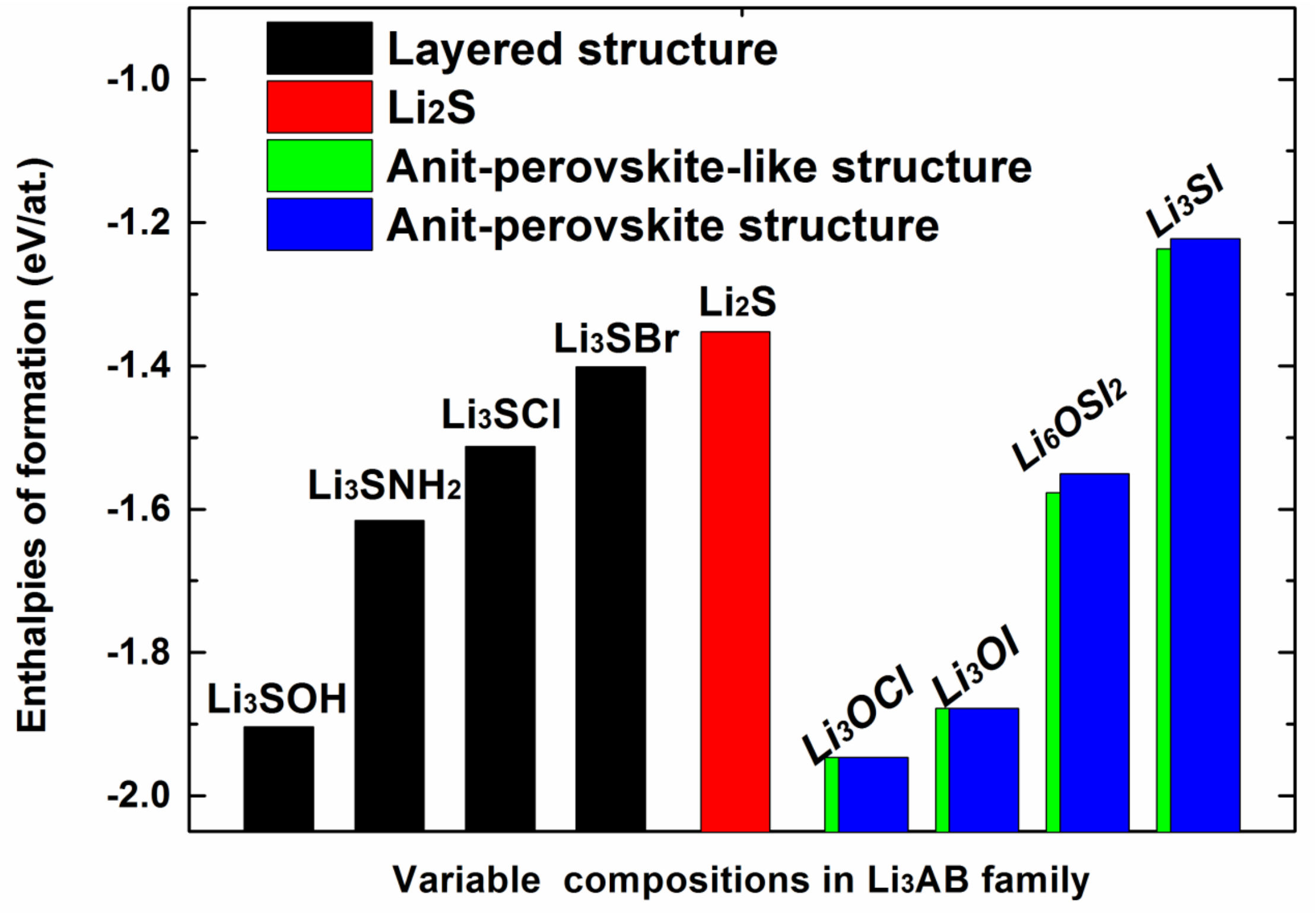

**Fig. 2** Associated fully-relaxed structures for (a) $Li_3OCl$, (b) $Li_3SCl$, (c) $Li_3SI$, and (d) $Li_6OSI_2$.

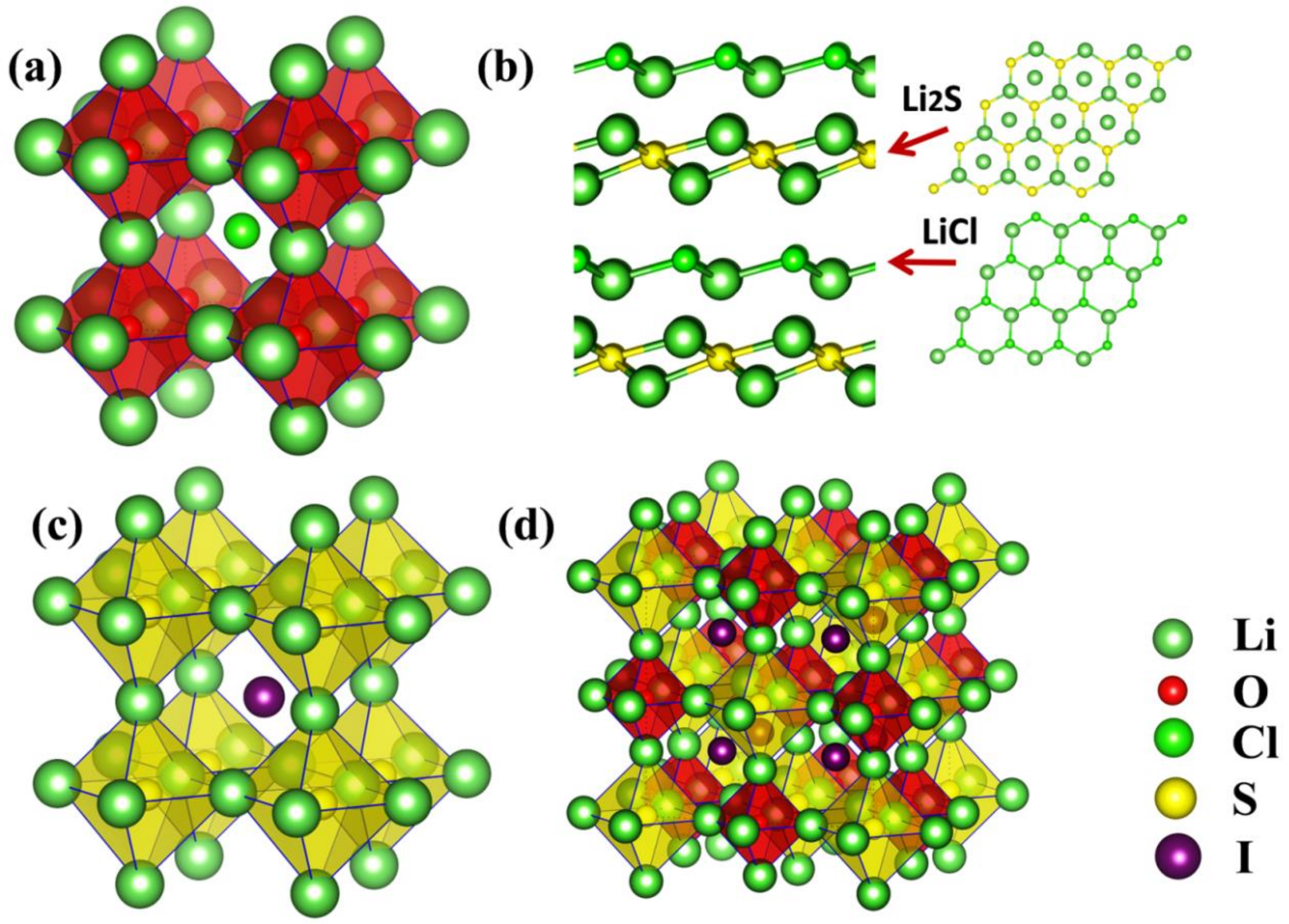

**Fig. 3** Calculated phonon band structures of the stable structures for (a) $Li_3OCl$, (b) $Li_3OI$, (c) $Li_3SI$ and (d) $Li_6OSI_2$.

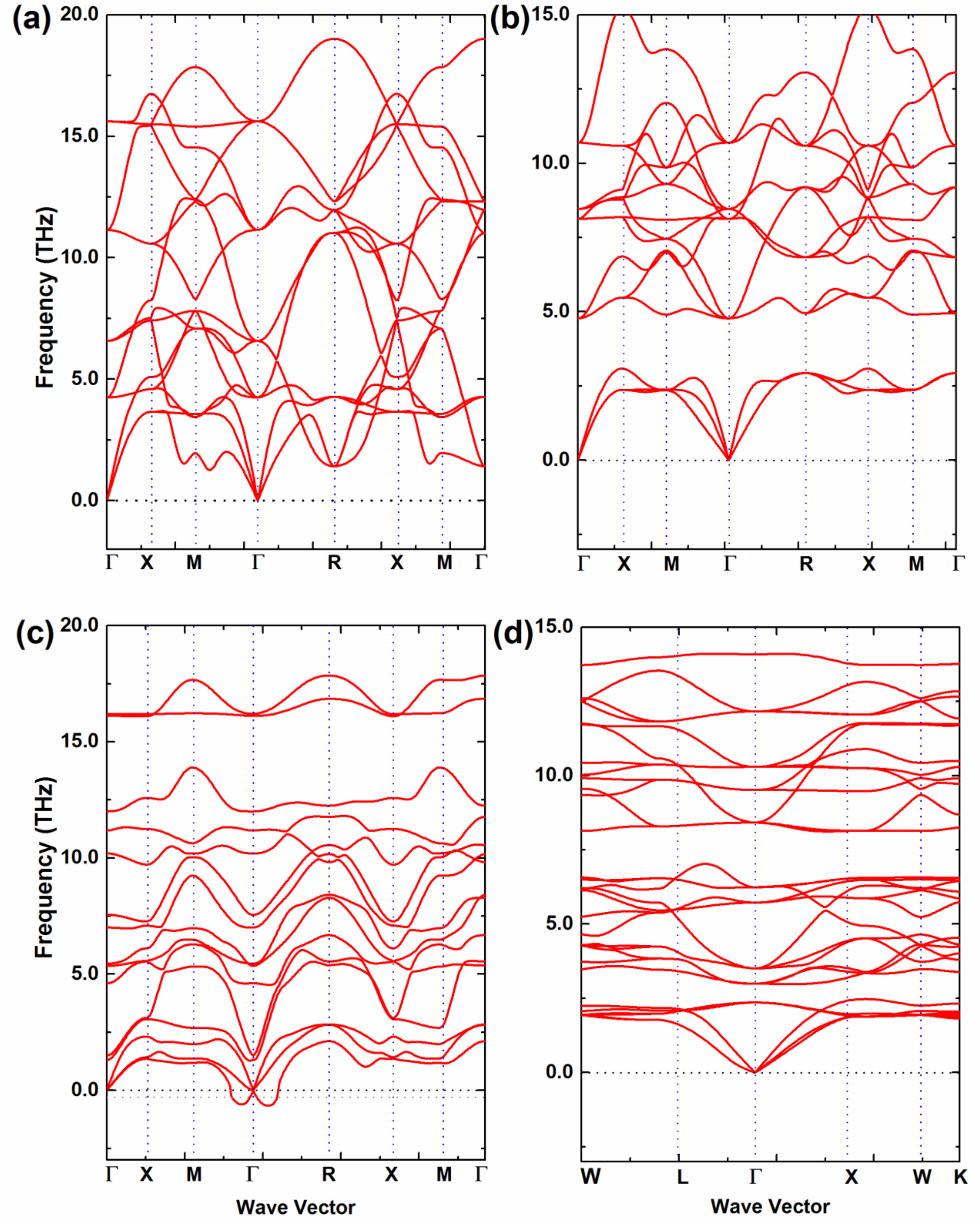

**Fig. 4** Phase diagrams for (a) $Li_3OCl$, (b) $Li_3SI$, and (d) $Li_6OSI_2$. Black circles represent stable compounds, and colored circles refer to metastable phases with energy of Hull being positive.

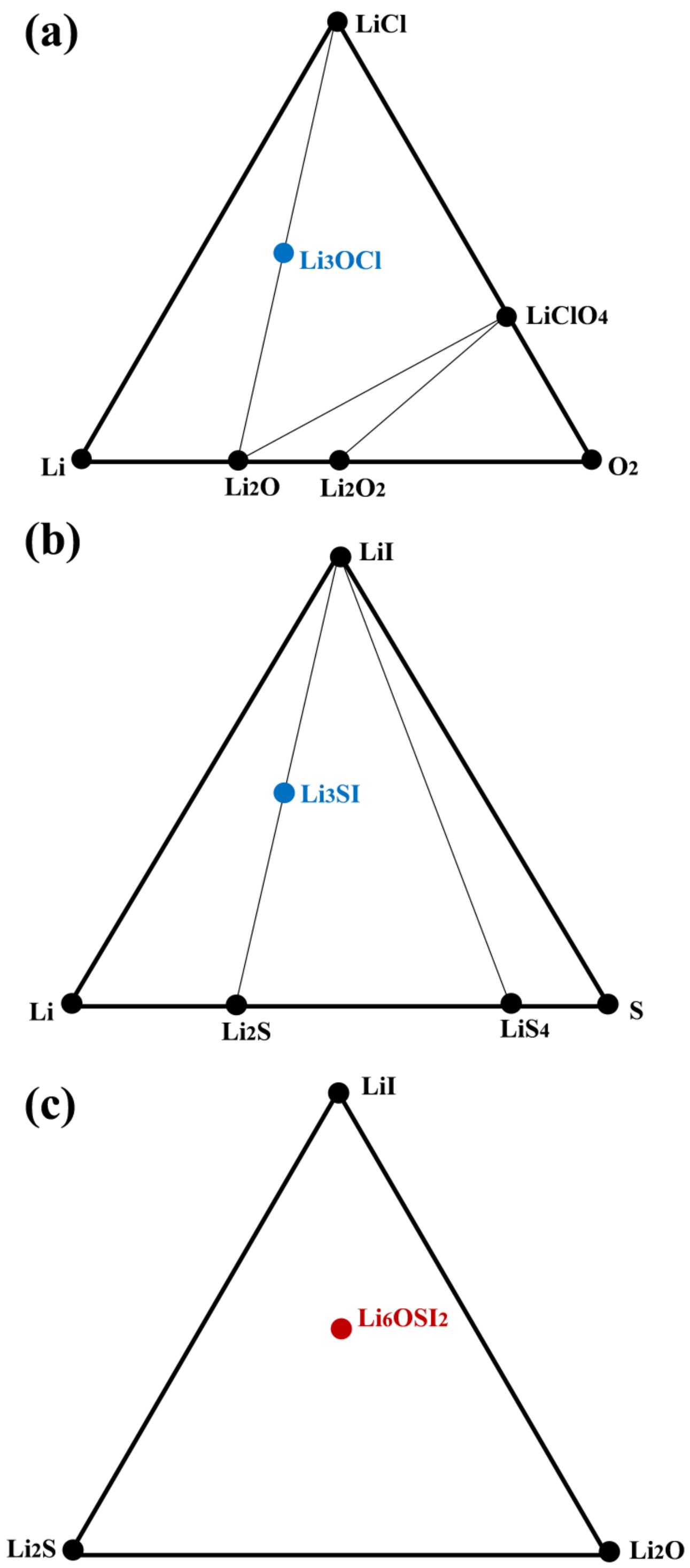

**Fig. 5** Free energy of formation per unit cell (uc.) for the compounds of $Li_3OCl$, $Li_3SI$, and $Li_6OSI_2$.

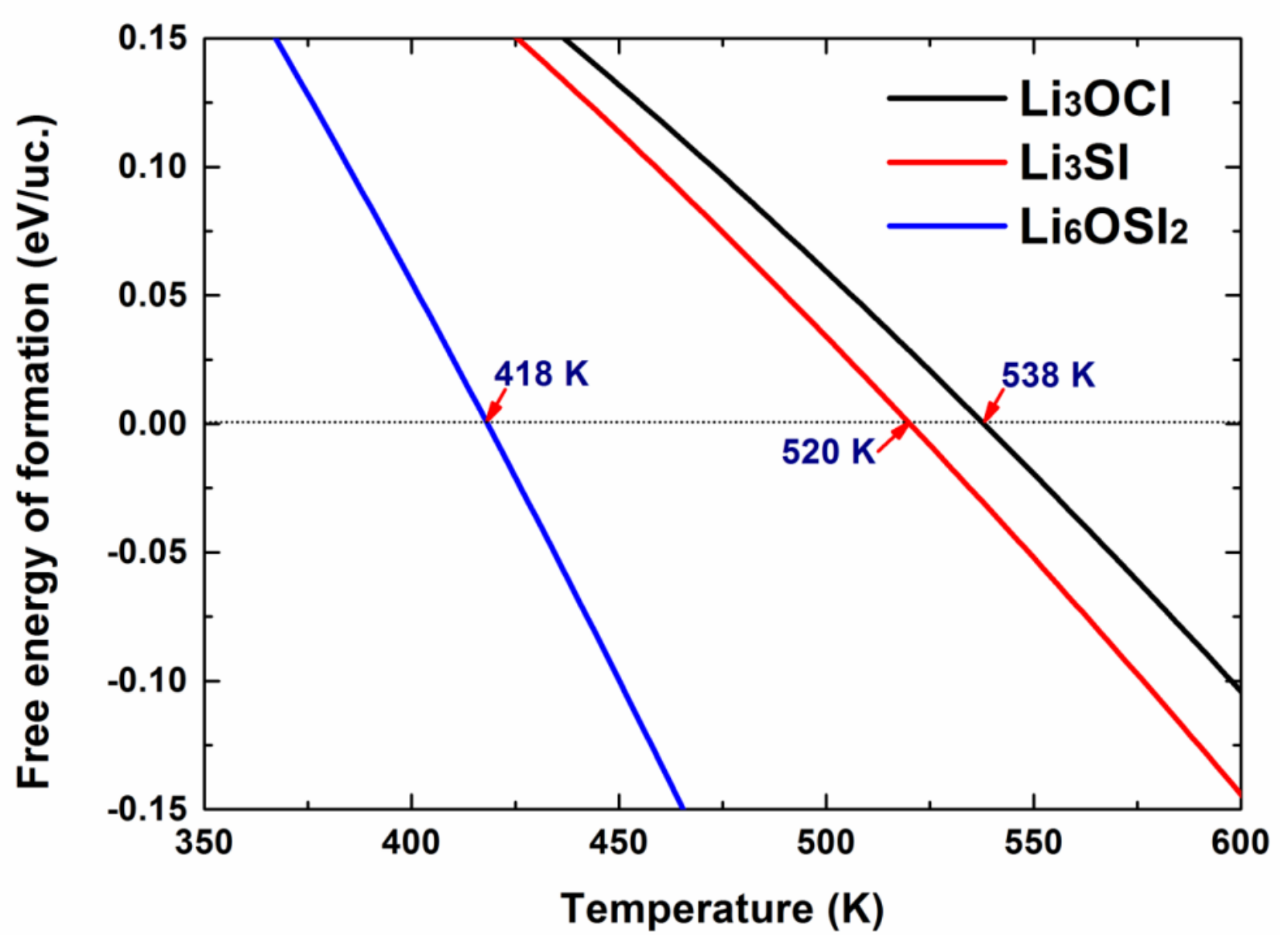

**Fig. 6** Plot of Li uptake per formula unit of solid electrolyte (red solid) against voltage vs Li/Li$^+$ for (a) $Li_3OI$, and (b) $Li_3SI$. Text indicates the predicted phase equilibria at corresponding regions at 0 K.

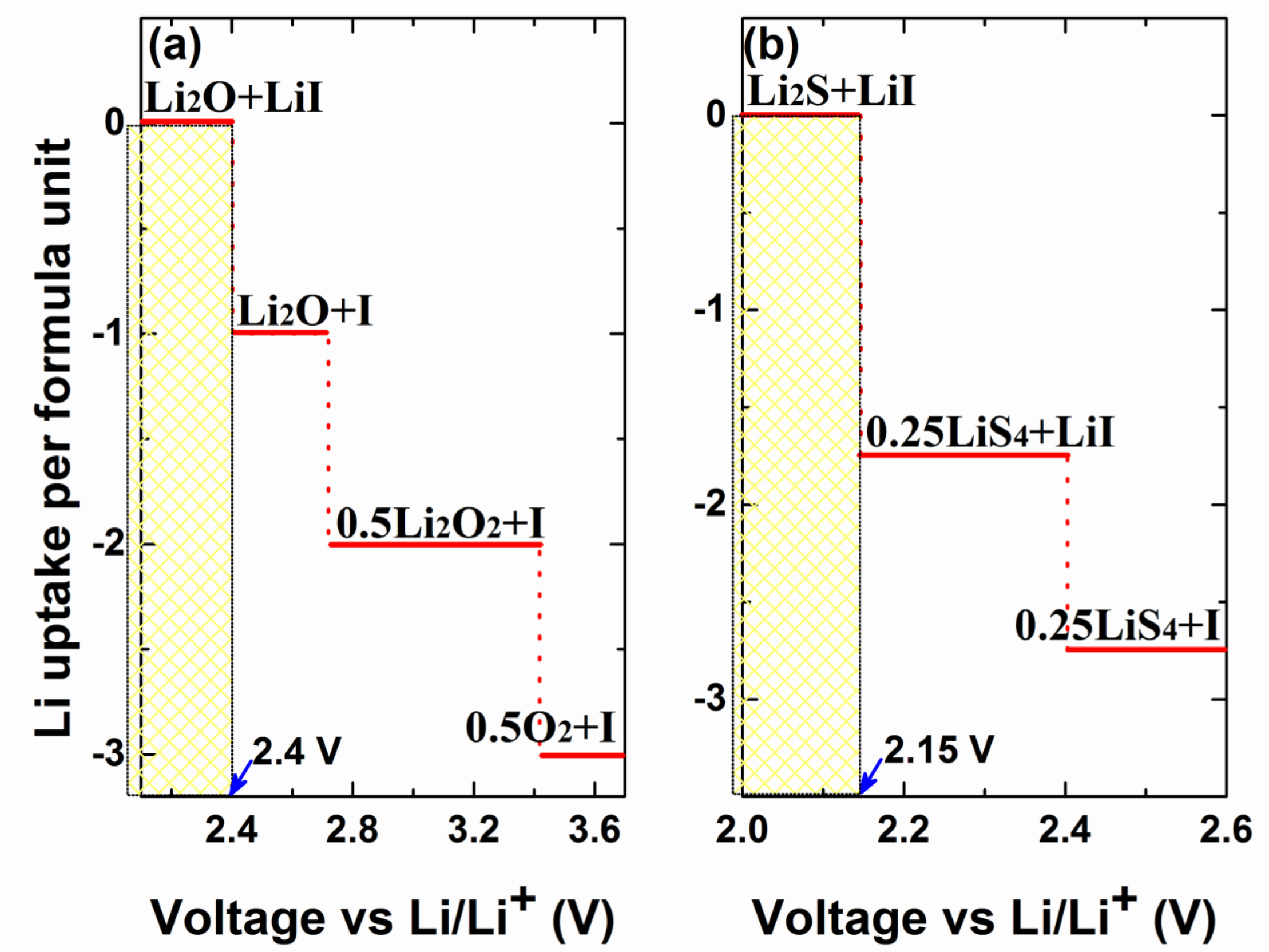

**Fig. 7** Calculated energy barriers for the migration of $Li^+$ ions in $Li_3SI$ along (a) $Li^+$ vacancy diffusive pathway, (b) the interstitial channel in a dumbbell pathway.

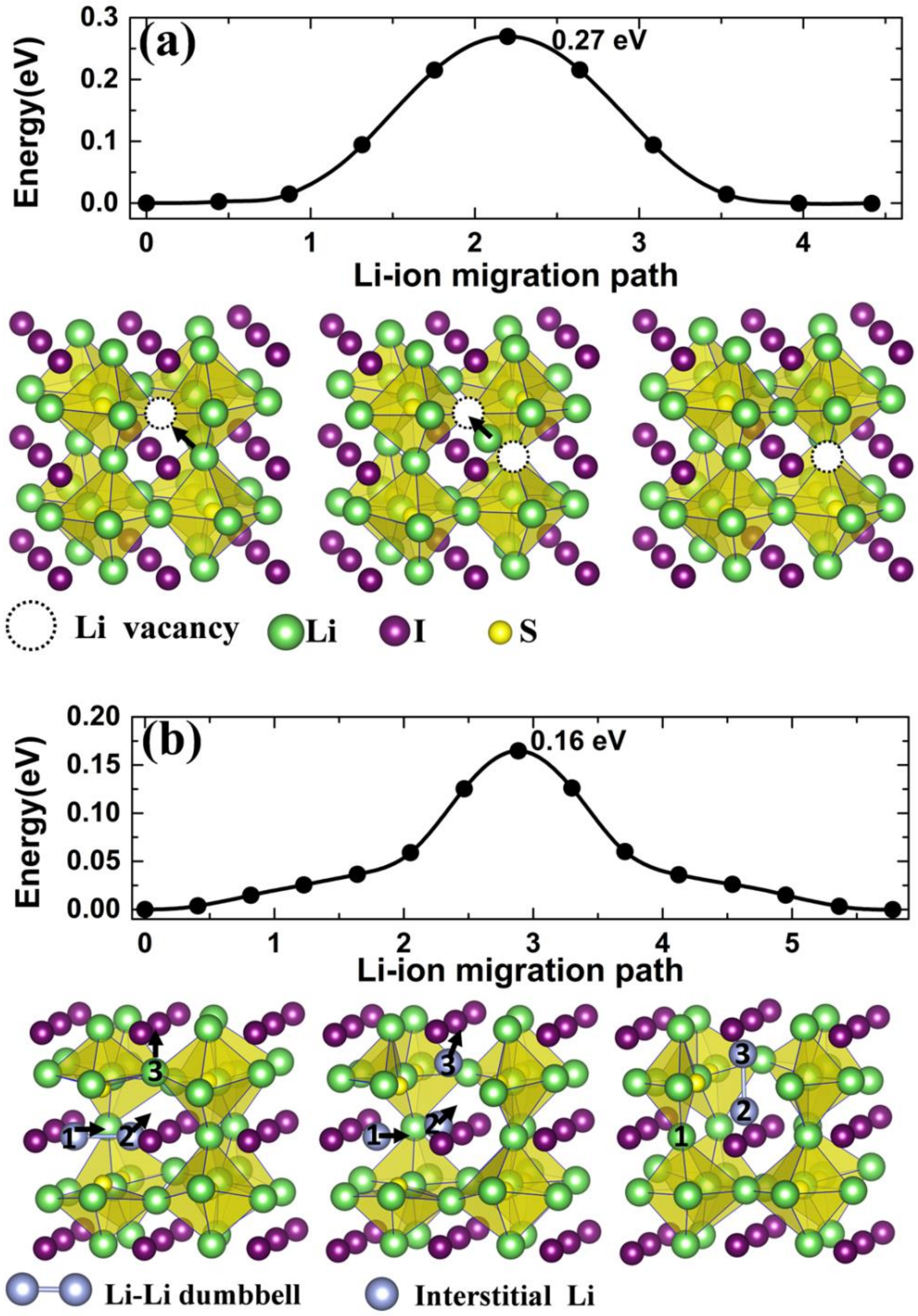

**Fig. 8** Charge distributions for the mid-path section of vacancy diffusive pathway for (a) $Li_3OCl$, (b) $Li_3SI$. And the interstitial channel for (c) $Li_3OCl$, (d) $Li_3SI$.

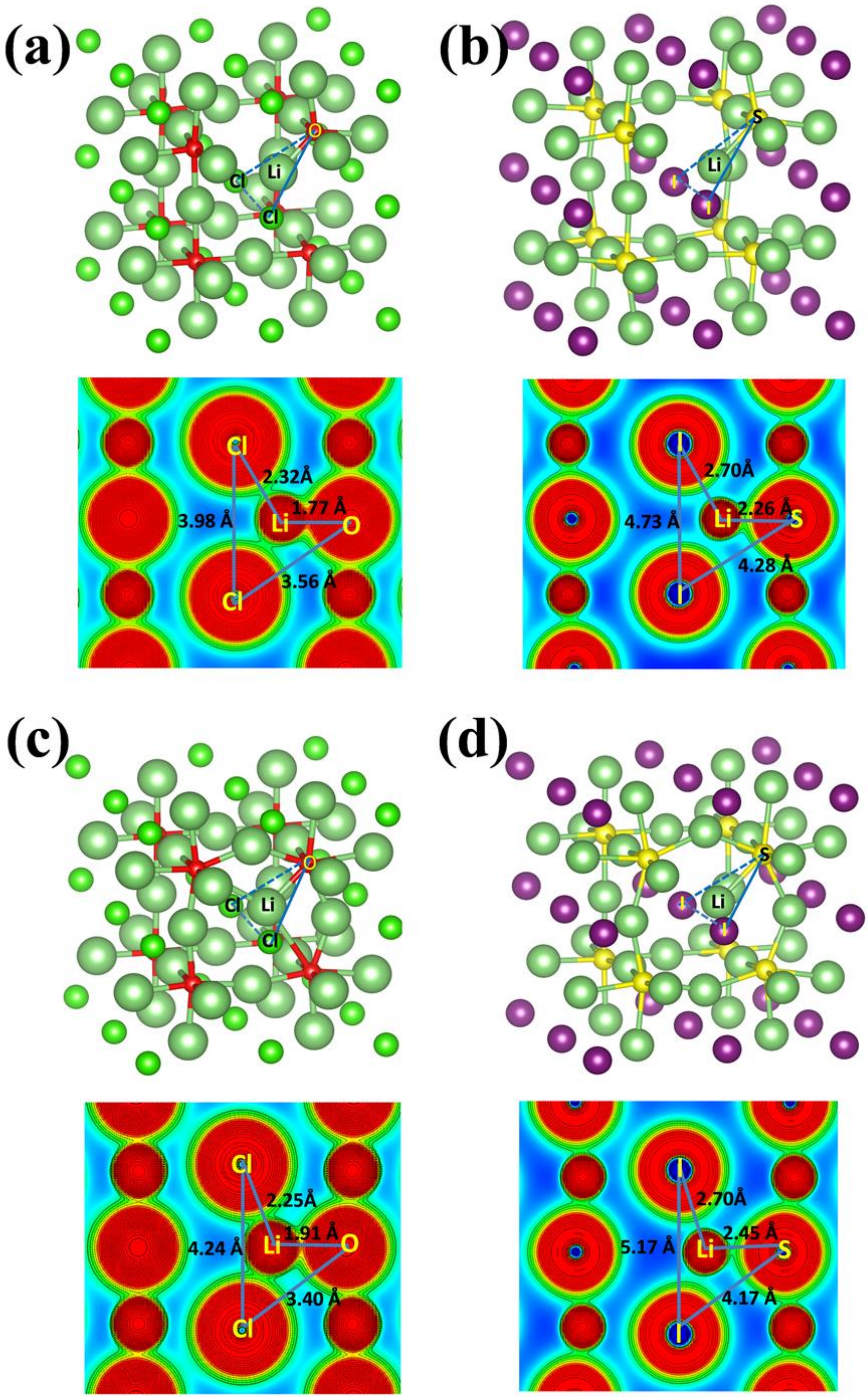

**Fig. 9 (a)** Two choices for the transportation of dumbbell Li+ ions in the $Li_6OSI_2$ phase, from 1-2 to either 3-4 or 3'-4'; (b) Calculated energy barriers for the migration of $Li^+$ ions around the $Li_6S$ octahedron and $Li_6O$ octahedron respectively.

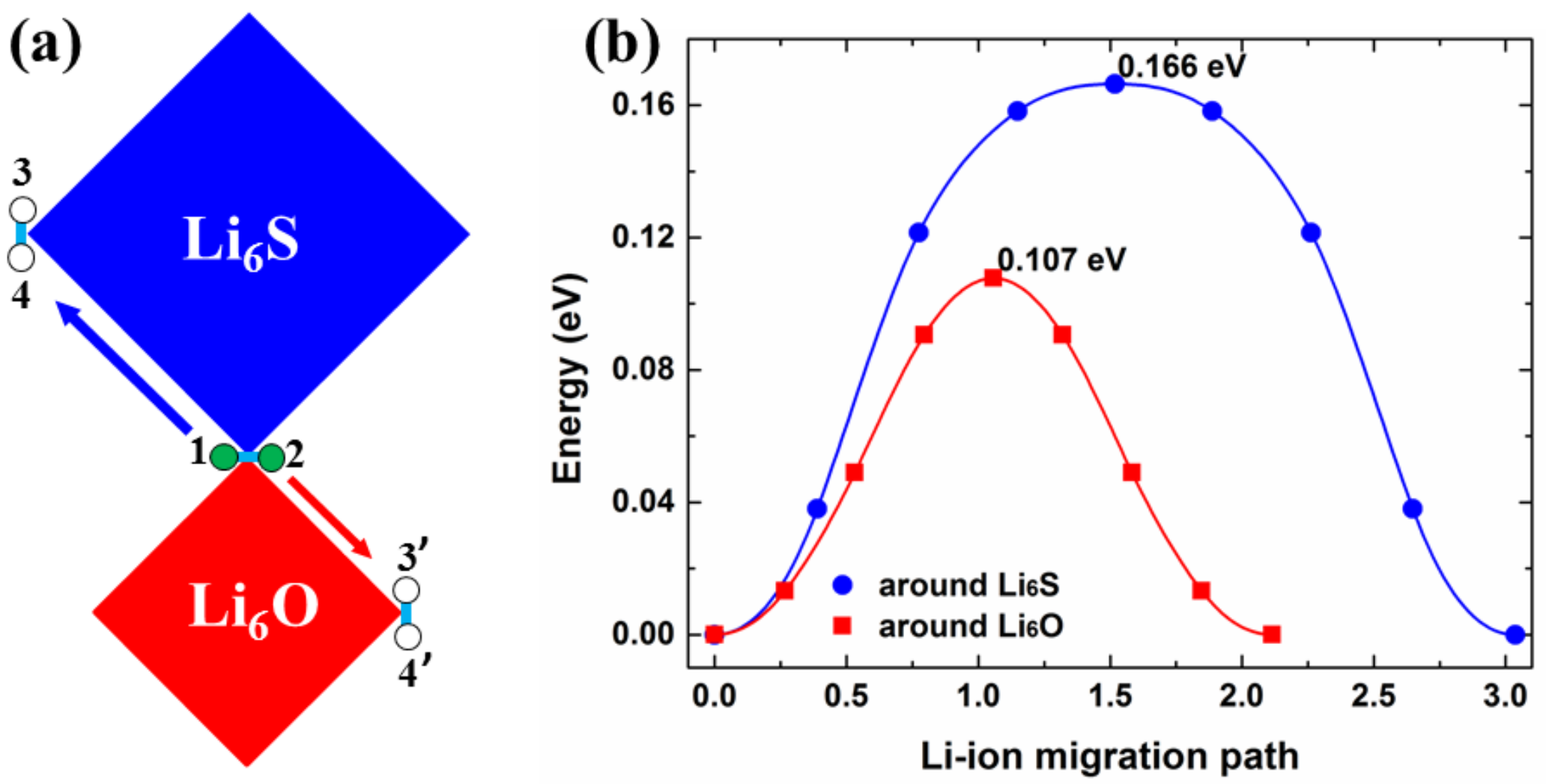

**Fig. 10** Diffusion coefficient for lithium ions from AIMD simulation. The D values at room temperature are presented by the hollow patterns.

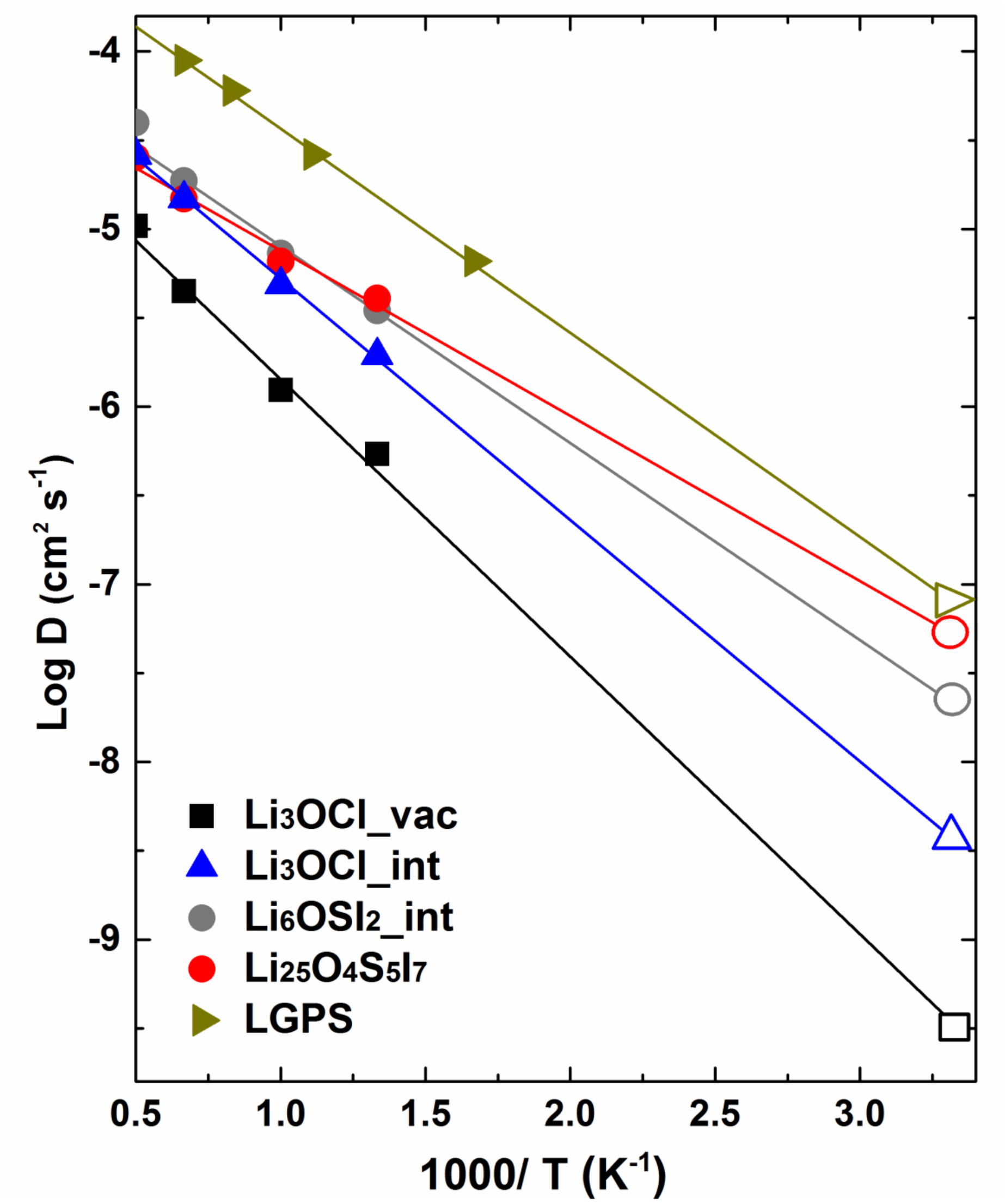

**Fig. 11** The projected density of states calculated using the HSE06 functional: (a) $Li_3OCl$, (b) $Li_3OI$, (c) $Li_3SI$, and (d) $Li_6OSI_2$.

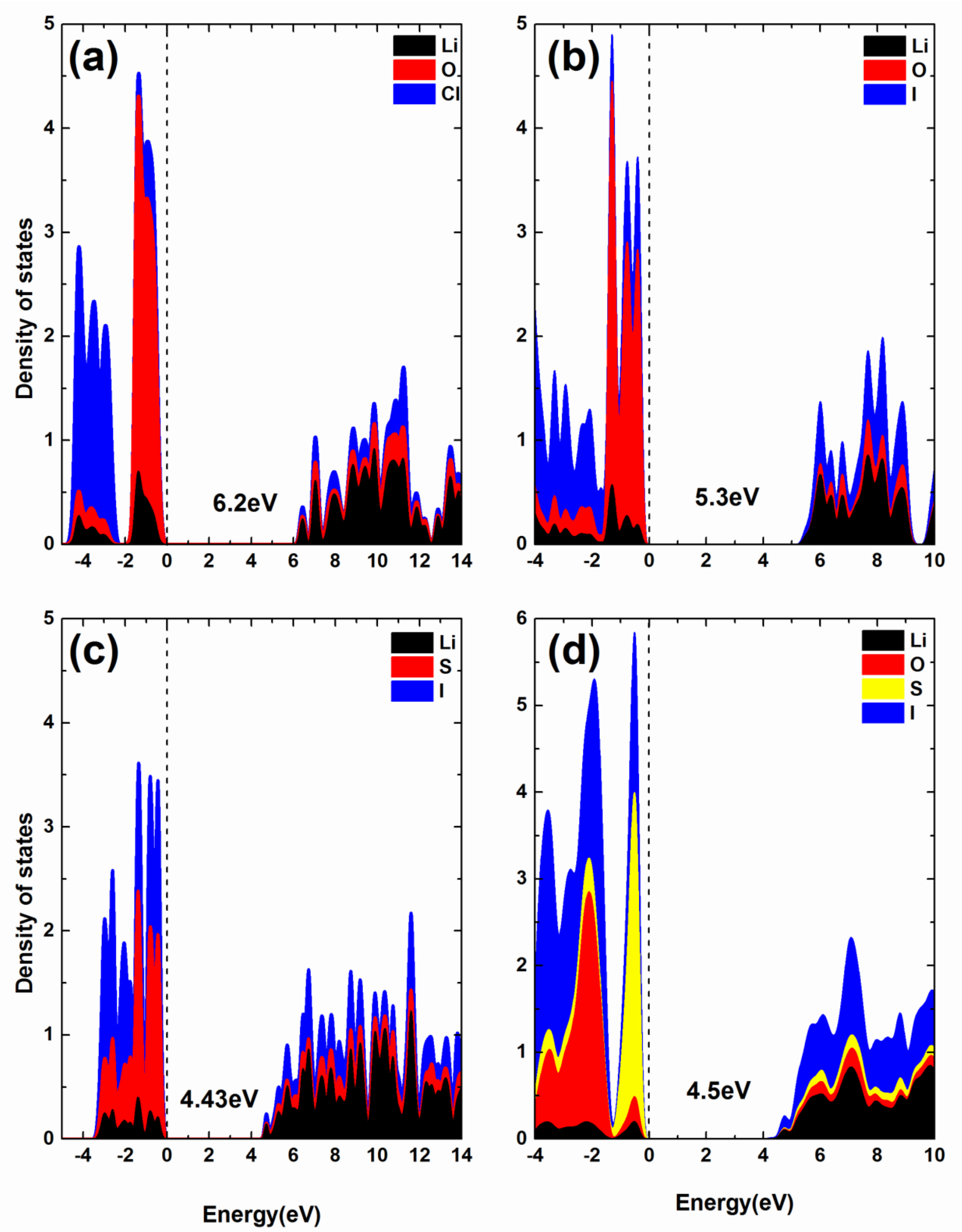

**Supplementary**

Materials synthesis:

$Li_6OSI_2$ solid electrolyte was prepared by high-energy ball milling and subsequent pressing without heating. $Li_2O$ (99.9% metals basis), $Li_2S$ (99.9%), and LiI (99.9% metals basis) powders were used as starting materials, which were mixed in the molar ratio $Li_2O:Li_2S:LiI = 1:1:2$ in an argon-filled glovebox ($O_2<0.1$ppm, $H_2O<0.1$ppm). The mixed powder was then transferred into an argon-filled zirconia milling container (100ml in volume), with the weight ratio of zirconia balls to powder being 8:1. Milling was carried out using a high-energy planetary ball mill at 280 rpm for 10h at room temperature. Crystalline powder of $Li_6OSI_2$ was obtained directly through cold pressing up to a pressure of 10Mpa in glovebox filled with argon.

**Fig s1** Powder XRD pattern for the cold-pressed $Li_6OSI_2$ sample with double-anti-perovskite structure (a), in comparison to calculated pattern from simulated structure (b). Trivial remnant LiI peaks are present, with peak intensities for the major phase being in excellent agreement with the theoretical pattern. Background hillock at lower angle range was from the encapsulating polymer for sample protection from the atmosphere.

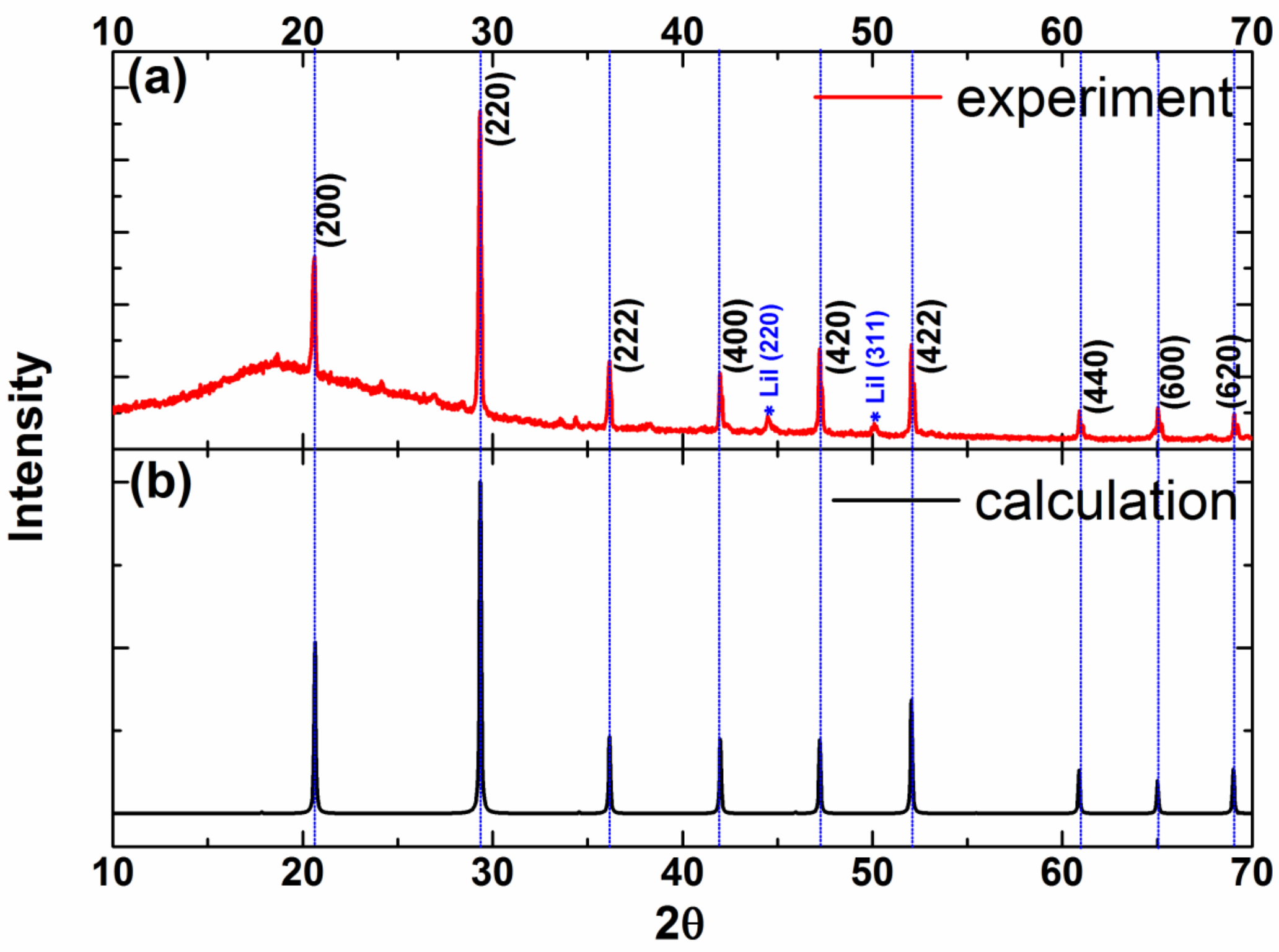

**Fig s2** Projected density of states calculated using the HSE06 functional for $Li_{25}O_4S_5I_7$

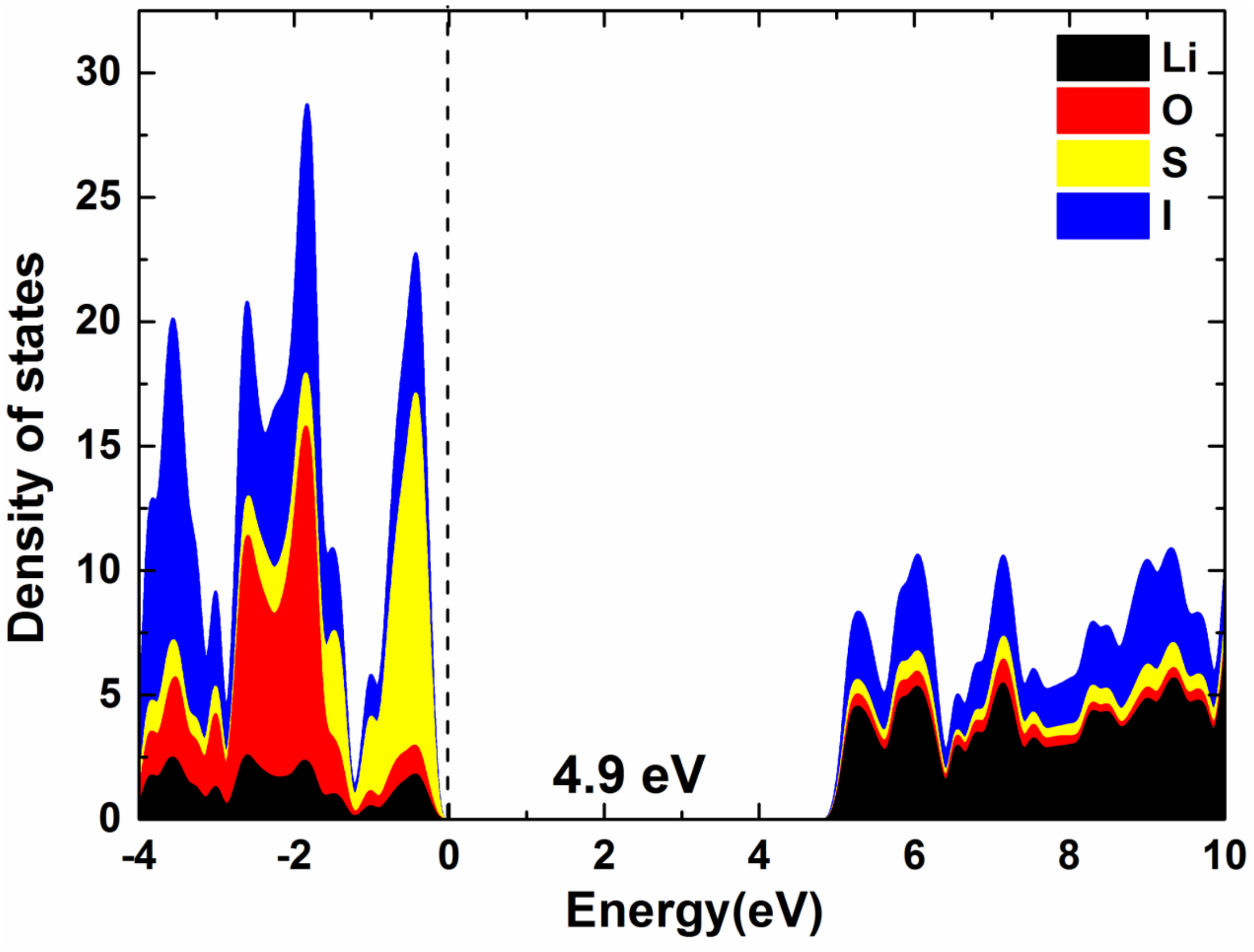